\documentclass[twocolumn]{revtex4}

\usepackage[pdftex]{graphicx}
\usepackage{amsmath}
\usepackage{amssymb}
\usepackage{marginnote}
\usepackage{verbatim}
\usepackage{array}
\usepackage{tabularx}
\usepackage{multirow}
\usepackage{color}
\usepackage{rotating}
\usepackage{color}
\usepackage[left]{lineno}

\begin{document}

\author{Ryan C. Taylor${}^{1,2,*}$, Xiaofan Liang${}^{2,*}$, Manfred D. Laubichler${}^{3,4}$, Geoffrey B. West${}^{5}$, Christopher P. Kempes${}^{5}$, and Marion Dumas${}^{6}$ \\
\footnotesize{${}^{1}$ School of Sustainability, Arizona State University, Tempe, AZ\\
${}^{2}$ Minerva Schools at KGI \\
${}^{3}$ Global Biosocial Complexity Initiative, Arizona State University, Tempe, AZ \\
${}^{4}$ School of Life Sciences, Arizona State University, Tempe, AZ \\
${}^{5}$ The Santa Fe Institute, Santa Fe, NM \\
${}^{6}$ London School of Economics and Political Science, London, UK \\
${}^*$Contributed equally\\
Correspondence and requests for materials should be addressed to C.P.K. (email: ckempes@gmail.com) and M.D. (email: m.dumas@lse.ac.uk)}
} 
\title{The Scalability, Efficiency and Complexity of Universities and Colleges: A New Lens for Assessing the Higher Educational System}

\begin{abstract} 
\noindent {\bf Abstract} The growing need for affordable and accessible higher education is a major global challenge for the 21st century. Consequently, there is a need to develop a deeper understanding of the functionality and taxonomy of universities and colleges and, in particular, how their various characteristics change with size. Scaling has been a powerful tool for revealing systematic regularities in systems across a range of topics from physics and biology to cities, and for understanding the underlying principles of their organization and growth. Here, we apply this framework to institutions of higher learning in the United States and show that, like organisms, ecosystems and cities, they scale in a surprisingly systematic fashion following simple power law behavior. We analyze the entire spectrum encompassing 5,802 institutions ranging from large research universities to small professional schools, organized in seven commonly used sectors, which reveal distinct regimes of institutional scaling behavior. Metrics include variation in expenditures, revenues, graduation rates and estimated economic added value, expressed as functions of total enrollment, our fundamental measure of size. Our results quantify how each regime of institution leverages specific economies of scale to address distinct priorities. Taken together, the scaling of features within a sector and shifts in scaling across sectors implies that there are generic mechanisms and constraints shared by all sectors which lead to tradeoffs between their different societal functions and roles. We particularly highlight the strong complementarity between public and private research universities, and community and state colleges, four sectors that display superlinear returns to scale.
\end{abstract}

\maketitle

\noindent {\bf Introduction}\\
How we deliver higher education globally and at increased rates to a growing population is a key challenge facing most societies today that will only increase in the future. It is not simply a challenge about growth, but equally about diversification and adaptation. In addition, universities and colleges play a central role in the future of human societies as complex social and environmental challenges require educated and active citizens, and as an increasing number of people need advanced training to participate in the ``knowledge economy'', including retraining in response to rapid technological innovations that shift the labor market \cite{RichardsTerkanian2013, WEF2018,CrowDabars2015}. These trends affect both mature and growing economies, as we see governments in countries with lesser penetration of higher-education set aggressive targets to catch up with more advanced economies (e.g. India's skill-building challenges and targets \cite{NSDC2015}), and both high-income and low-income economies struggling to finance education \cite{Rose2005,Ferrantietal2003,SchmidtTraubSachs2015}. In addition, the details of institutional design and educational strategy matter for the ultimate success of graduates \cite{Brookings2015,CrowDabars2015} and, by implication, of societies.  

Universities and colleges in the United States represent one of the most diverse instances of a higher education system in that they span sizes from 10 to over 200,000 enrolled students, contain both nonprofit and for-profit models, and range in strategy from vocational training to the production of novel research \cite{Clark1986}. Consequently, analyzing the US higher education system can provide important insights about how the basic mechanisms, tradeoffs, and outcomes of university function relate to size as societies aim to scale-up total educational outputs. To date, our understanding of these tradeoffs and capabilities as a function of institutional structure, educational mission, and overall size is still limited and important policy decisions have to be made without sufficient empirical evidence.  

Scaling laws have provided important insights across the entire spectrum of science and technology ranging from understanding the fundamental forces and constituents of nature to the dynamics and structure of human engineered systems, biological organisms, and social organisations \cite{brown,west-gene,west1997general,kempes,kempes2016evolutionary,kempes-2,bettencourt2007growth,bettencourt2010urban,bettencourt2013origins,Bettencourtetal2014,vanRaan2013,Jamtveitetal2009,Mathiesenetal2010}. A number of studies have shown how systematic analyses of scale can reveal underlying mechanisms \cite{brown,west-gene,west1997general,kempes,kempes2016evolutionary,kempes-2}. As yet, no such analysis has been done for universities and colleges, which make up the higher education sector. In biology, a longstanding theory of organism scaling has motivated extensive empirical work, finding that many physiological and life-history characteristics ranging from metabolism and growth rates to life expectancy vary systematically with body mass \cite{brown,west1997general,west2002}. Theoretical advances on the origins of these phenomena have led to predictions of universal biological behavior, biogeography, evolutionary transitions, growth dynamics, and detailed physiological tradeoffs \cite{brown,west-gene,west1997general,kempes,kempes2016evolutionary,kempes-2}. The application of scaling theory to social systems has also revealed important regularities: for example, measures of human creativity increase predictably with city size, with the super-additivity of human interactions in social networks being the driving mechanism \cite{bettencourt2007growth,bettencourt2010urban,bettencourt2013origins,Bettencourtetal2014,vanRaan2013,Jamtveitetal2009,Mathiesenetal2010}. This literature illustrates that the scaling perspective can effectively (i) illuminate key systematic behavior and tradeoffs, (ii) define the most appropriate way of standardizing features by the size of the system (for example showing when per capita measures are inappropriate in social systems), (iii) identify fundamental mechanisms and constraints, and (iv) make predictions. 

These scaling relationships typically take the simple mathematical form of power laws:
\begin{equation} 
\label{eq1}
Y=aX^{\alpha}
\end{equation}
where $Y$ is a property of interest in the system, $X$ is the size of the system, $\alpha$ is the scaling exponent, and $a$ is a normalization constant. For instance, in cities, data show that almost all socio-economic metrics, from total wages and GDP to the number of social interactions and number of patents produced, scale with population size as a power law with an exponent of $\approx 1.15$. This is an example of what is commonly referred to as superlinear scaling (exponent larger than 1). Consequently, on a per capital basis, socio-economic metrics increase proportionally to $X^{0.15}$, implying that on a per capita basis, larger cities promote more social interactions and greater production of patents, and therefore more innovation \cite{HendersonCockburn1996,bettencourt2007growth,bettencourt2010urban}.

 Analyzing scaling relations in universities provides us with a fascinating case study for applying scaling theory in that universities are a class of entities that share a subset of overlapping goals, but also manifest radically different strategies and fine-grained differences in institutional objectives. Furthermore, many universities are currently undergoing rapid transformations which may be expressed as changes in overall scaling relationships due to shifts in their internal structures. This can potentially provide a diagnostic tool for understanding the mechanisms underlying long-term trends in the performance of these institutions with applications for designing higher education.

Our findings include: first, that universities do indeed exhibit scaling behavior, and that the seven commonly used sectors for characterising institutional differences -- research universities (public and private), state colleges, community colleges, non-profit private colleges, for-profit colleges, and professional schools -- follow very different scaling regimes. For example, consider research universities. We find that they scale superlinearly in revenues and expenditures (i.e. these variables grow faster than linearly with the size of the institution). They diversify into more activities with size, accrue prestige and wealth, becoming increasingly active in research but expensive for students. In contrast, we find that revenues and expenditures in state and community colleges scale with exponents that are less than 1, that is, they scale sublinearly: increasing size allows them to decrease costs to students and taxpayers faster than linearly.  Second, we find that almost all of these sectors display similar economies of scale in one or more components of their expenditure streams, particularly in instruction costs, but also in maintenance and bureaucratic costs. Third, we observe that universities in different groups leverage these economies of scale in different ways, which support different goals, ranging from expanding research, to increasing access to education, or increasing profits. Fourth, we discuss the tradeoffs between the different functions of universities which could explain these patterns, thereby providing a synoptic view of how different types of universities differ in their ability to further these functions at scale.

As a consequence, this novel perspective into the entire higher education system, which reveals broad systematic quantitative insights into its structure and taxonomy, provides a fundamentally new framework for fruitfully informing policy-makers to respond to the challenging needs of society. Furthermore, it also provides university and college administrators a new, potentially powerful, tool for understanding the stated roles of their institution and for assessing its performance  relative to other institutions.

\noindent {\bf Scaling Framework}

In this paper we describe the scaling behavior of basic processes in universities: their inputs, including revenue, faculty and students, and their outputs, including expenditures, graduation rates and other related outcomes that fulfill key societal purposes. Our focus on scaling is close to the economists' approach to measuring economies of scale and scope in multi-product firms \cite{Baumoletal1982}, but is less parametric and more general in that we do not presuppose the form of the cost function. Instead, we consider different variables' scaling relationship individually, and, then, by contrasting them and taking a synoptic view, highlight the salient variations and properties of the institutions. In contrast to the economics approach where student enrollment is, in turn, taken as an input or an output \cite{DeWitteLopezTorres2017}, depending on the model, here we consider student enrollment to be a fundamental property of the system, treat it as the independent variable, and ask whether it systematically structures the institution. One advantage of first analyzing variation with scale, is that we can more easily identify whether the differences in two universities' mix of inputs and outputs is a result of their size difference, or whether it reflects different management strategies. From a scaling perspective, we focus mostly on the value of the exponent, $\alpha$, which leads to the classification of systems as follows:
\begin{enumerate}
\item $\alpha > 1$  : superlinear scaling; this points to increasing returns to scale (if $Y$ is an output), or diseconomies of scale (if $Y$ is an input).
\item $\alpha = 1$  : linear scaling; this points to constant returns to scale (if $Y$ is an output), or constant economies of scale (if $Y$ is an input).
\item $\alpha < 1$  : sublinear scaling; this points to decreasing returns to scale (if $Y$ is an output), or economies of scale (if $Y$ is an input).
\end{enumerate}
If $X$ is an input in the production of $Y$, $Y/X$ gives the average cost, which is proportional to $X^{\alpha-1}$. If $\alpha$ is less than one, the unit cost is decreasing with system size, indicating economies of scale. 
\vspace{2mm}

\noindent {\bf Expected scaling in higher education} \\
The scaling approach is well positioned to enrich the study of organizations. The guiding question is that of scalability: what, if anything, limits the size of firms, institutions, and societies \cite{Coase1937,JohnsonEarle2000,Axtell2001}? What tradeoffs between multiple productive and bureaucratic functions accompany growth? To this end, scaling supplies a natural quantitative connection to structuralist theory of organizations in economics, sociology and anthropology. Universities provide a unique class of institutions to test how differences in internal strategy alter overall scaling relationships, which has applications not only for designing higher education but understanding how the mechanisms behind social scaling could be adjusted.

How should we expect scale to affect the internal processes of universities? Past efforts have found varying patterns of economies of scale and scope, but more consistently that universities tend to operate near their optimal size and surprisingly near their efficiency frontier \cite{DeGrootetal1991,Cohnetal1989,Izadietal2002,Johnesetal2005}. In contrast, the broader higher-education literature has tended to cast doubt on the efficiencies of universities, with little regard for size. It has focused instead on the alarming rising costs of education (average full-time student tuition in the U.S. increased by 113\% from 1984 to 2014 \cite{NCESquick}). Papers and reports in the higher-education literature have suggested the causes are increases in the wages of professionals \cite{ArchibaldFeldman2008}, changing market structure \cite{Hoxby1997}, but also increase in non-instructional professional services and associated administrative costs \cite{Clotfelter2014,LeslieRhoades1995}. In this literature, some fault universities for their profligate spending, accusing them of excessively diversifying into non-core activities, while others point to personalized attention and diverse campus activities as a key to success after graduation \cite{WebberEhrenberg2010}. Others point to the pernicious effects of the race for prestige amongst the top-tier universities \cite{CrowDabars2015,BokHigherEd}. 

Here we distinguish between two main processes: production processes -- teaching and research -- and maintenance processes of administration and operations. Teaching is the most fundamental production process. Teaching expenditure is dominated by the remuneration of academic staff, itself the product of the number of faculty and the mean faculty salary. Scale can thus impact teaching expenditure via either of these variables. As a university increases in size, it has the possibility of exploiting economies of scale in the number of faculty by increasing class sizes. Universities may follow this strategy, possibly at the risk of compromising educational quality and outcomes. Research is another important production process for the subclass of universities that engage in it. Research, very much like new patent production in cities, is a creative process, which is typically assumed to be driven by the frequency and diversity of social interactions.  We expect this to scale superlinearly with the number of university researchers and enrollment, similar to the scaling of patent production in cities with increasing population size \cite{bettencourt2010urban}. \cite{vanRaan2013} provides indirect evidence for universities, documenting that citations scale superlinearly with the size of research universities. For research universities, the increased research activity that we expect to see with increased size could have both positive and negative effects on student learning: it gives students access to research staff, but may draw resources away from teaching. It should also be noted that students not only affect increasing returns by supporting a larger faculty and campus, but are also themselves an input to the education system through peer learning and to the research enterprise as participants. 

Maintenance processes include all aspects of administration and institutional support. An important general hypothesis is that larger organizations let bureaucracy grow out of proportion because they differentiate into a wider range of operations \cite{Blau1970} and must monitor more personnel \cite{RasmusenZenger1990,McAfeeMcMillan1995}. This mechanism would put a limit on the size of organizations. It has been suggested that the growing size and complexity of social organizations lead to a disproportionate growth in maintenance processes, portending the collapse of entire societies \cite{tainter1988collapse}. At the same time, the economies of scale of each operation would seem to promise unbounded growth \cite{Blau1970}. 

All of these hypotheses are of interest, which we explore in different university sectors using 2013 data from the Delta Cost Project (here ``delta data'') \cite{DeltaDoc2017} (see methods).

\begin{figure*}
\centering
\includegraphics[width = .8\linewidth]{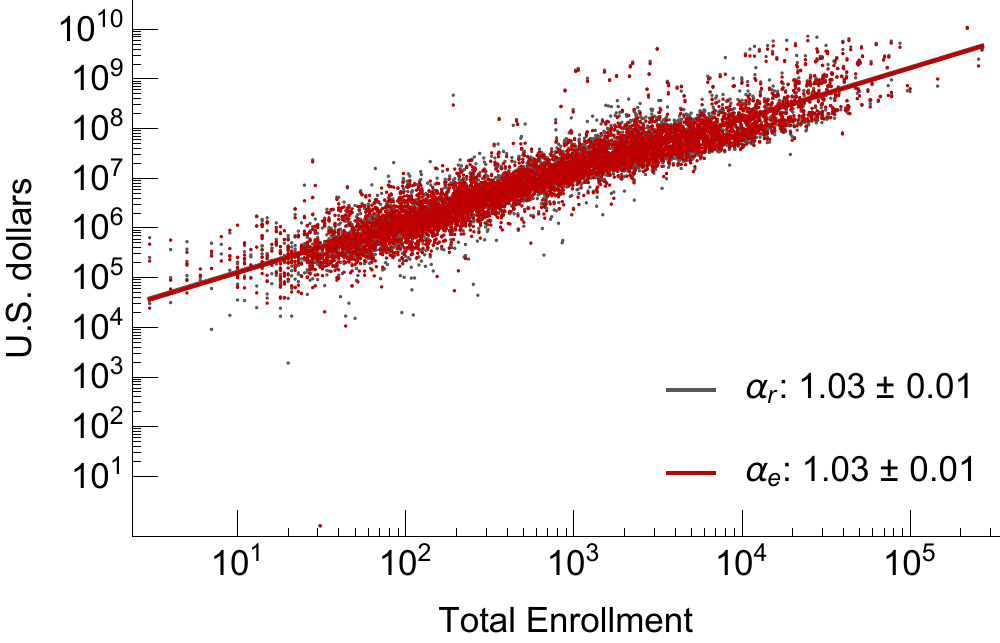}
\caption{The scaling relationships between total revenue (subscript ``r'') and student enrollment, and total expenditure (subscript ``e'') and student enrollment, combining all schools in the dataset. Note that revenue and expenditure are generally very well matched so both the data points and the regression lines overlap, which explains why much of the revenue data and the revenue regression line are hidden under the expenditure data and regression line.}
\label{aggregatescaling}
\end{figure*}

\noindent {\bf Results}\\
\noindent {\bf Scaling in Higher Education}\\
In all of our analyses we use total student enrollment as the natural measure of size as we are ultimately interested in the resources and benefits provided to the individual student (see SI Figure C1 for alternatives). Figure \ref{aggregatescaling} shows the total financial throughput (total expenditure and total revenue) of all universities and colleges, pooled together regardless of their sector, plotted as a function of their size. This clearly demonstrates that, as a totality, they do indeed systematically scale with size, strongly supporting the use of scaling as a methodology for revealing underlying regularities and mechanisms common across all universities and colleges. The figure shows that financial throughput scales linearly with size, suggesting that, on average, there is no advantage to being larger at least as far as these economic indicators are concerned. However, this masks significant underlying diversity of behavior between different educational sectors, arising from the wide diversity of mission and strategy amongst universities. 

\begin{figure*}
\centering
\includegraphics[width = 1.0\linewidth]{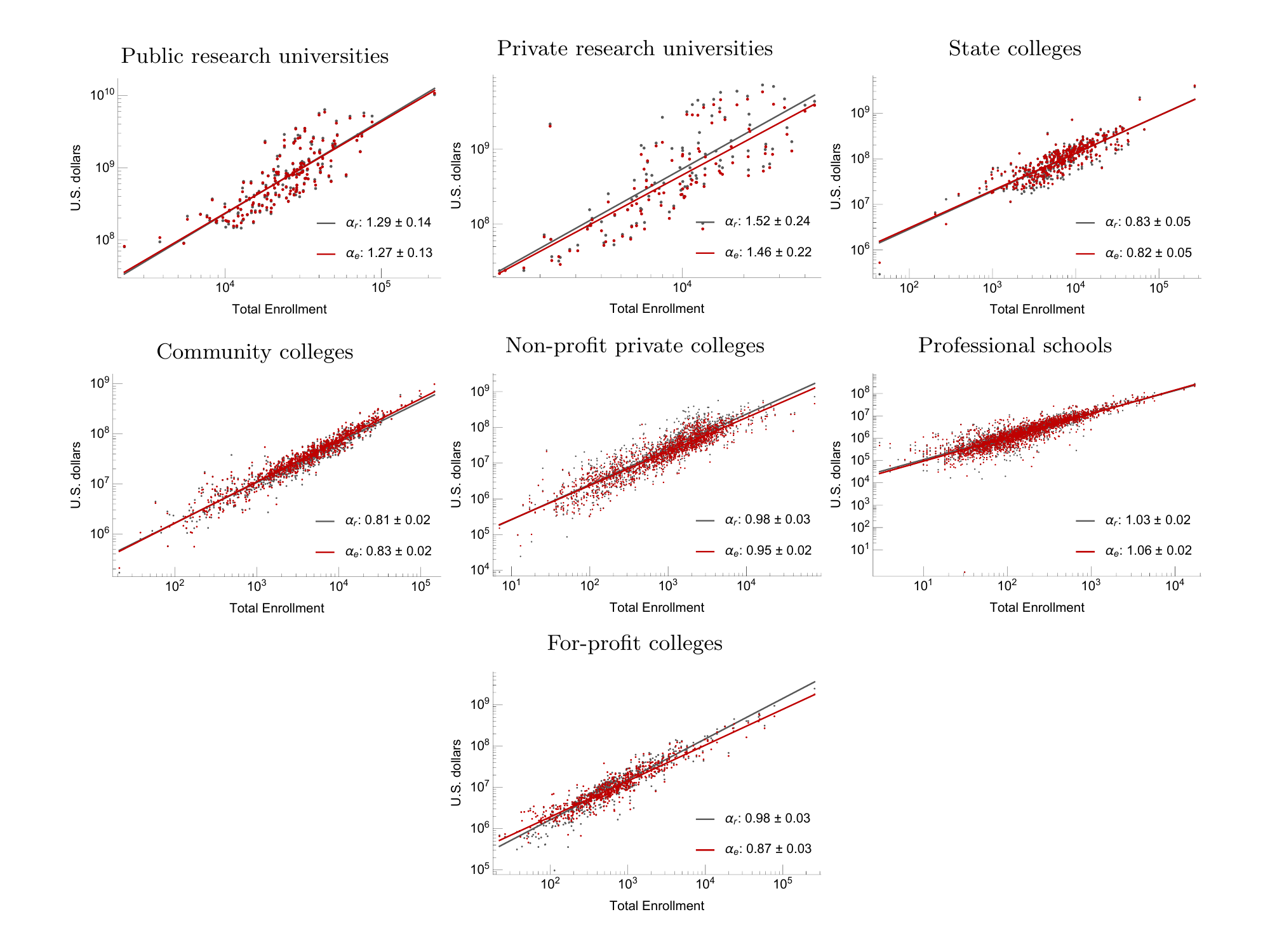}
\caption{The scaling of total revenue (subscript ``r") and total expenditure (subscript ``e'') as a function of total enrollment by sector. The regression lines may overlap, with expenditure hiding the revenue regression line.}
\label{fig:revexp}
\end{figure*}

To see this, we classify all universities and colleges into seven conventional sectors according to institutional control, level, and research activity (outlined in Table 1; see SI Tables D1-D2 and Figure D1 for detail). As shown in Figure \ref{fig:revexp}, these different sectors display dramatically different scaling behaviors for total revenue and expenditure. At this level of granularity, we can distinguish between four broad regimes. First, research universities (both public and non-profit private) scale superlinearly: as they enroll more students, their revenues and expenditures increase faster than linearly, in other words, financial throughput per student increases with size (note however the large confidence interval for the private research sector). Second, and in marked contrast, community colleges and state colleges display remarkably sublinear scaling, that is, financial throughput per student decreases with size, representing strong overall economies of scale. Third, non-profit private colleges and professional schools scale roughly linearly with size, indicating little advantage in being larger. Fourth, for-profit colleges display linear scaling in revenue but sublinear scaling in expenditure, which implies that they are able to make a profit by exploiting economies of scale in their costs. 

\begin{figure*}
\centering
\includegraphics[width = 1.0\linewidth]{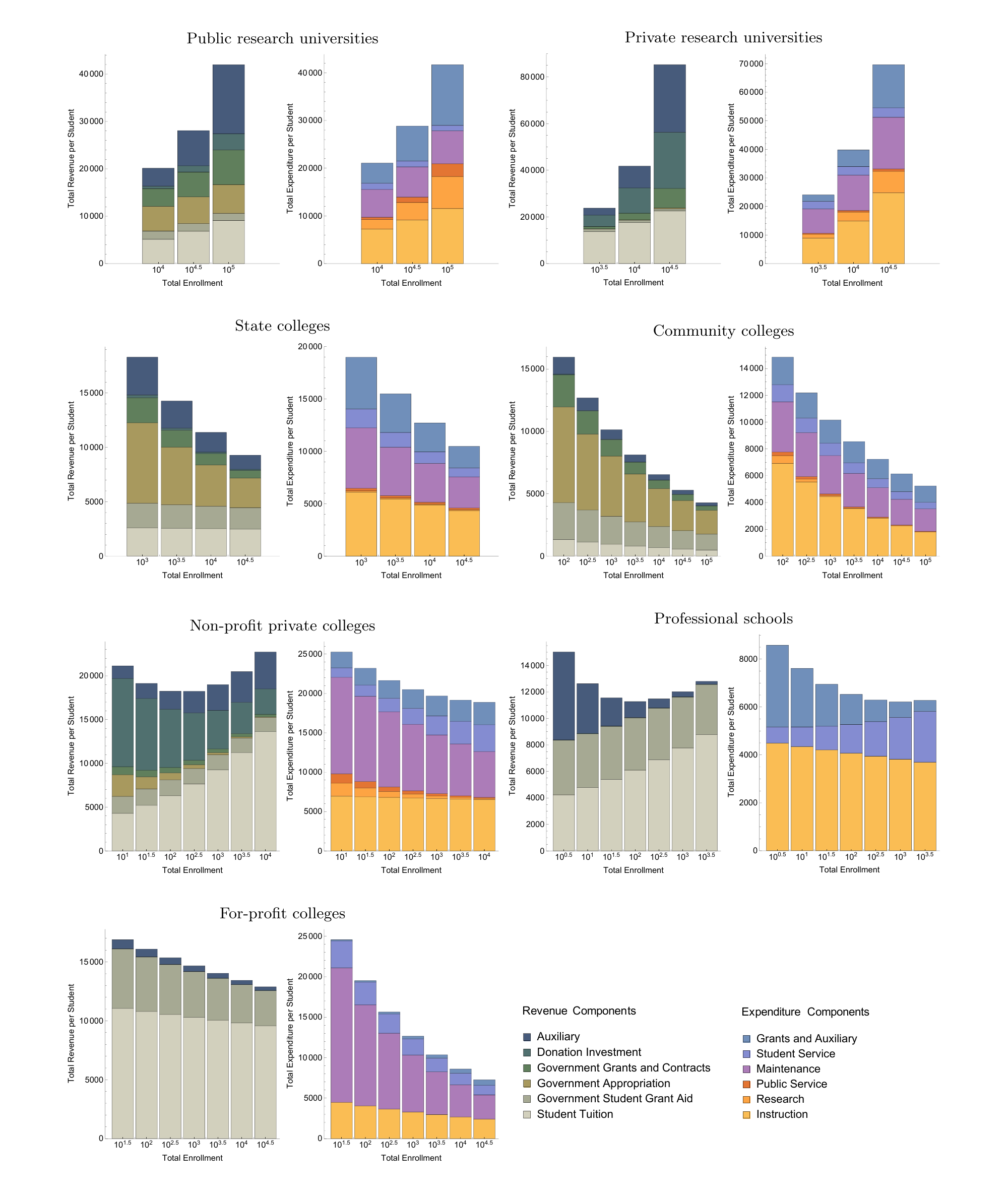}
\caption{Variation in the internal composition of revenue and expenditure, shown per student and as a function of institution size. This stacked representation makes clear that different sectors display dramatically different total economic streams. Note that an increase in the total height of a bar with institution size indicates superlinear scaling, and a decrease  indicates sublinear scaling. See SI Figure J1 for the regression coefficients underlying each plot.}
\label{subcomponents}
\end{figure*}

To better understand the different strategies of these sectors, Figure \ref{subcomponents} shows the detailed scaling of components of revenues and expenditures, which allows us to examine the relative importance of various university activities with changes in university size (see SI Figures F1-F4 for explanation of how these plots are constructed). Table 2  summarizes the most salient of these activities (teaching expenditure, research expenditure, tuition revenue, and maintenance) along with the scaling of several other key educational inputs and outcomes (faculty size, research revenue, student completions, and student mid-career earnings). Typically, we find that as universities grow, they find specific areas of economies of scale that they then exploit to further their core mission. At the same time, Table 2 shows clearly that different sectors differ markedly in the areas in which they display respectively superlinear, linear or sublinear scaling, suggesting that there are tradeoffs between the different functions universities choose to play. The table also suggests that we can summarize the typology of universities according to their scaling behavior into four distinct regimes:  

\begin{enumerate}
\item Research universities (public and private) scale superlinearly in all activities and sources of revenue, but sacrifice affordability. As they grow larger, they seek to attract increasingly prestigious faculty (as indicated by the superlinear scaling in faculty pay, especially in private universities) and charge higher tuition, also attracting better resourced students, who later on enjoy higher earnings. The fact that both research and educational outcomes scale superlinearly suggest that these activities are synergistic.
\item State and community colleges display very strong sublinear scaling in teaching expenditure and total faculty. This translates to some extent into sublinear scaling in tuition revenue and potentially compensates for the observed sublinear scaling in public funding revenue. Their baseline graduation rates are low compared to research universities, but stay constant or increase with the size of the school despite lower costs. Hence, for the same likelihood of achieving a degree, they become increasingly affordable with size, either to students, or taxpayers, or both.
\item Non-profit private colleges and professional schools expand student services disproportionately with increasing size, and come to rely increasingly on tuition revenue. Tuition scales superlinearly, while graduation rates scale only linearly. Therefore, they become less affordable with size for a similar probability of graduating. 
\item For-profit colleges display strong economies of scale in all areas of expenditure, but tuition revenue scales linearly, which implies that they become increasingly profitable with size. Unfortunately, we do not have data on student completions.
\end{enumerate}

These regime characteristics suggest that research universities on the one hand, and state and community colleges on the other, display particularly favorable scaling relationships, but in non-overlapping functions. These four sectors (and two regimes) therefore seem strongly complementary, fulfilling different societal functions. We will come back to this in the Discussion. We now provide a detailed analysis of each sector, in which we examine their distinct economic strategies and how it relates to the outcomes in Table 2. 
\vspace{2mm}

\noindent {\bf Public research universities}\\
In line with the expectations we outlined earlier in the paper, research activities (as measured by research expenditures) and research output (proxied by revenue from grants) scale superlinearly with size in public research universities (Figure \ref{pubresRE}). We note, however, that the proxy we use for research output is not very precise. We also look at data on research funding provided by the NSF and find this relationship to be very uncertain (see SI Figures F5-F6 and Table F1). Along with this, teaching expenditure also scales superlinearly with size: the amount of money spent per student increases more than proportionally with the number of students. Thus, far from exploiting the potential economies of scale in teaching, public research universities pursue an opposite strategy: they increase both the student-to-faculty ratio and the salaries of their faculty. Indeed, as Table 2 shows, this trend in instruction expenditure is the combination of a superlinear increase in the total number of faculty (dominated by full-time faculty), and a moderate increase in the average salary of faculty members (in Table 
2, "Faculty Pay" is the total sum paid to all faculty so that the average faculty salary increases if and only if the exponent for "Faculty Pay" is higher than the exponent for "Total Faculty"). For example, in a university of $5,000$ students, the faculty-to-student ratio is 9:100, with faculty paid on average $\$ 42,500$/yr, while in a university of $50,000$ students, the faculty-to-student ratio is 13:100, with faculty paid on average $\$ 47,000$/yr. This suggests an interesting interaction between research and instruction as universities grow in size: as research becomes more rewarding and important, universities seek to attract a greater number of professors (as well as graduate students), competing more fiercely for sought-out faculty, thereby raising faculty pay. 

Maintenance and administrative costs scale slightly superlinearly, but do not systematically outpace production processes (teaching, research and student completions). This indicates that there are no apparent diseconomies of scale in maintenance function. On the contrary, efficiency in maintenance seems to support the diversification of activities in line with the hypothesis in \cite{Blau1970}.

Given the superlinear increase in instruction expenditure, it is perhaps not surprising that the number of students completing their degree scales superlinearly with the size of the student cohort. In particular, we note the very high scaling exponents for first-time full-time student completion ($1.24$ for public universities). This superlinear scaling in completions is accompanied by a superlinear increase in tuition, indicating that these schools attract better resourced students \cite{Clotfelter1999}. For example, a school of $5,000$ students will typically charge $\$6,650$ with a $63\%$ graduation rate (using FSA completion rates), while a school of $50,000$ students will typically charge $\$10,100$ with an $78\%$ graduation rate. Consistent with this result, the number of students receiving federal financial aid (FSA students) scales sublinearly (see SI Appendix G). While the completion rates amongst these students scales superlinearly, it does so more weakly and in absolute terms is lower than for first-time full-time students, suggesting that the socio-economic background of students plays an important role in explaining outcomes in these schools. 
\vspace{2mm}

\noindent {\bf Private research universities}\\
Private non-profit research universities behave similarly to public ones with a few critical distinctions. First, we note that this sector displays a lot more variability than other sectors (see confidence intervals in Table 2). This is likely due to a greater variety of institutional models, with some institutions running very large non-degree granting government and private research centers. Second, for any size, tuition is much higher than in the case of public research universities, and also scales with a higher exponent. The scaling behaviors of expenditure and revenue streams are dominated by the disproportionate increase in instruction expenditure and tuition revenue, respectively. The data on faculty numbers and faculty pay also reveal an interesting difference. In private research universities, the superlinear scaling in faculty pay betrays an important increase in average faculty salary with school size (from an average of $\$ 48,700$/yr in a school of $5,000$ students to $\$81,000$/yr in a school of $50,000$ students). Despite these differences between the private and public sectors, the superlinear scaling of completions is not significantly different.

Our analysis suggests that research and education act synergistically since student outcomes increase with increases in research, which is consistent with prior findings on educational economies of scope \cite{Johnesetal2005}.
The data also suggests that as research universities grow larger, they become more prestigious, more successful, but also more expensive for students. At the larger end, the public and private universities' pattern of expenditure and revenue become very much alike, with large public research universities attracting private money in addition to public funding, and private research universities attracting federal appropriations in addition to private funds. 
\vspace{2mm}

\noindent {\bf State colleges}\\
State colleges stand in stark contrast to research universities. First, they display very strong economies of scale in instruction, largely due to sublinear scaling of the number of faculty, thus decreasing faculty-to-student ratio as schools increase in size (Table 2). For example, a state college of $1,000$ students has a faculty-student ratio of 8:100, whereas a school of $50,000$ students has a faculty-student ratio of 5:100. Faculty salaries, on the other hand, scale significantly higher than faculty number, so each instructor earns systematically higher wages at larger schools. Nonetheless, total faculty pay exhibits economies of scale ($\alpha= 0.91$). We also see very strong economies of scale in maintenance costs (with a scaling exponent $\alpha = 0.80$). Other areas of expenditure (student service, auxiliary expenditure) also scale sublinearly. 

Surprisingly, this impressive decrease in per capita expenditure is accompanied by superlinear scaling in the completion rate of students, with a scaling exponent of $1.11$, both for students receiving financial aid and other first-year, first-time students. The completion rate for FSA students in state colleges is $47\%$ for a college of $1,000$ students but increases to $60\%$ for a college of $10,000$ students\footnote{This value does not include students who transferred to another institution} (compared with $68\%$ for a university of $10,000$ students in the public research sector, and $90\%$ in the private research sector). 
Yet, what our scaling analysis reveals is that larger schools do increasingly better, despite lower expenditures (in particular lower faculty numbers), the larger schools reaching graduation rates of $60\%$. Possible explanations are that students benefit from the increasing opportunities for social interactions in larger schools, that larger schools attract more applicants and are therefore more selective, or that larger schools offer a greater diversity of courses, better satisfying the demands of students, despite the larger class sizes. External factors could be at play, such as incentives to graduate arising from the local labor market. While often overshadowed by their public research counterparts, the state colleges fulfill an essential role in the American higher education system and seem to be particularly well positioned to provide higher-education at scale.

Another noteworthy feature of state colleges is that tuition scales linearly. Thus, the reduction in expenditure does not drive a commensurate decrease in tuition. This is even clearer if we replace total enrollment with the full-time equivalent number of students to account for part-time students, in which case we find that tuition increses slightly superlinear (see SI Appendix C). One reason is that  appropriations, as well as local grants, decrease significantly with the size of the university (Figure \ref{statecollRE}). Hence, at scale, state colleges educate more students at lesser cost to the taxpayer, and, overall, affordability for an equal probability of completion tends to be higher, or at least non-decreasing, for larger state colleges. 
\vspace{2mm}

\noindent {\bf Community colleges} \\
Community colleges behave similarly to state colleges, but display even more pronounced economies of scale. Expenditures in instruction decreases dramatically on a per capita basis. This is driven by a sublinear scaling of the number of faculty, while the average instructor salary remains constant. Maintenance and administration are also increasingly efficient, characterized by an exponent $\alpha=0.88$. 

A majority of students in community colleges do not complete their degrees (the average completion rate for FSA students is $30\%$). This is to be expected because this specific sector attracts substantial numbers of non-degree seeking students, and often caters to them. With this mind, it is striking that student completions in the public 2yr sector scale linearly with the size of the FSA cohort, which is evidence that larger community colleges at least maintain their capacity to retain students despite very large economies of scale in expenditures and increasing class sizes. Furthermore, once we consider the educational outcomes of students who transfer to a 4yr institution, we see a superlinear increase in the number of students securing a 4yr diploma. In other words, students at smaller colleges more often stay for Associate Degrees, while students at larger ones tend to secure a Bachelor's -- arguably a better educational outcome (see SI Appendix G).

In line with their public service mission, community colleges take advantage of their cost savings to reduce the cost to attending students. Indeed, tuition scales decisively sublinearly with total enrollment (with a fairly low exponent $\alpha = 0.89$), in stark contrast to research universities and non-profit private colleges. This seems partially due to the increase in the number of part-time students with scale, who pay lower tuition (see SI Appendix C). Per capita tuition revenue at a $5,000$ student community college is on average $\$2,200$, while it is $\$1,700$ at a college of $50,000$. Additionally, outside revenue from appropriations, donations and grants, also scale sublinearly, and even more dramatically than student revenue. These schools thus operate with a tighter and tighter budget at scale, providing education at a decreasing cost to the taxpayer.
\vspace{2mm}

\noindent {\bf Non-profit private colleges}\\
Non-profit private colleges (which include liberal arts colleges) behave very differently from research universities or public colleges (see SI Table D4 for results specific to liberal arts colleges). First, as with all the other sectors so far, they display economies of scale in maintenance and administration. In contrast, instruction expenditure remains constant on a per capita basis (Figure \ref{libartsRE}). Interestingly, this is due to the combination of sublinear scaling of the number of faculty (decreasing faculty-to-student ratio), combined with an increase in the average faculty salary. Hence non-profit private colleges, as they become larger, pay fewer but more expensive faculty, keeping their instruction expenditure per student constant. Meanwhile, we observe a marked increase in student services expenditures, a form of diversification of the school's activities with scale.

The graduation rate at these schools is fairly high (on average $55\%$) and remains the same for schools of different sizes (traditional completions appear slightly superlinear, see SI), suggesting no systematic changes to educational output with scale, despite dramatic increases in student services. Interestingly, donations, endowment revenue, and appropriations scale sublinearly. To finance the increase in student services despite this decrease in several revenue sources, these schools become increasingly focused on increasing tuition as they grow in size. Indeed, tuition scales strongly superlinearly ($\alpha = 1.15$). This indicates that affordability for an equal probability of completion decreases for larger schools in this sector. 
\vspace{2mm}

\noindent {\bf Professional schools}\\
In for-profit professional schools, expenditure scales slightly superlinearly. This increase is accounted for by a ramp up in student services, while instruction expenditure slightly decreases on a per capita basis, similar to the non-profit private colleges (Figure \ref{profschoolsRE}). This sector has the most drastic reduction in total faculty number with enrollment (with a scaling exponent $\alpha = 0.76$). 

Data on completion is very scant for professional schools. We only have data on first-time first-year student completions from the same two-year college within three years, which scale linearly ($\alpha=1.02$). These completions are paired with a slightly superlinear growth of total tuition revenue ($\alpha=1.09$). Indeed, without the support of any private investment, tuition quickly becomes the overwhelming source of funds for these schools. Notably, unlike for-profit colleges, schools in this sector do not seem to use economies of scale to increase their profit as the enrolled population grows. 
\vspace{2mm}

\noindent {\bf For-profit colleges}\\
As with state colleges, these schools are able to reduce their per capita instruction costs as the school grows in size. 
As in the case of state colleges and non-profit private colleges, this reduction in instruction costs with size can be decomposed into a decrease in the faculty-to-student ratio (note the very strongly sublinear exponent for total faculty $\alpha = 0.83$), combined with an increase in the average instructor salary. The sublinear scaling in instruction is paired with a strongly sublinear scaling in academic support and student services, in contrast to non-profit private colleges. Overall, this private for-profit sector displays dramatic expenditure reductions in all areas, on par only with state and community colleges. 

Neither traditional nor FSA graduation rate data were reliable enough for us to assess returns to educational outcomes with size. However, we can still assess affordability and profitability. Tuition scales linearly with both total and FTE enrollments, indicating consistent access to these colleges across the full range of their sizes. All other four-year sectors show higher scaling of tuition. However, Figure \ref{forprof} shows that the difference between revenue and expenditure systematically widens with scale, which indicates that this sector uses economies of scale to increase profitability. 

\begin{figure*}
\centering
\includegraphics[width = 1.0\linewidth]{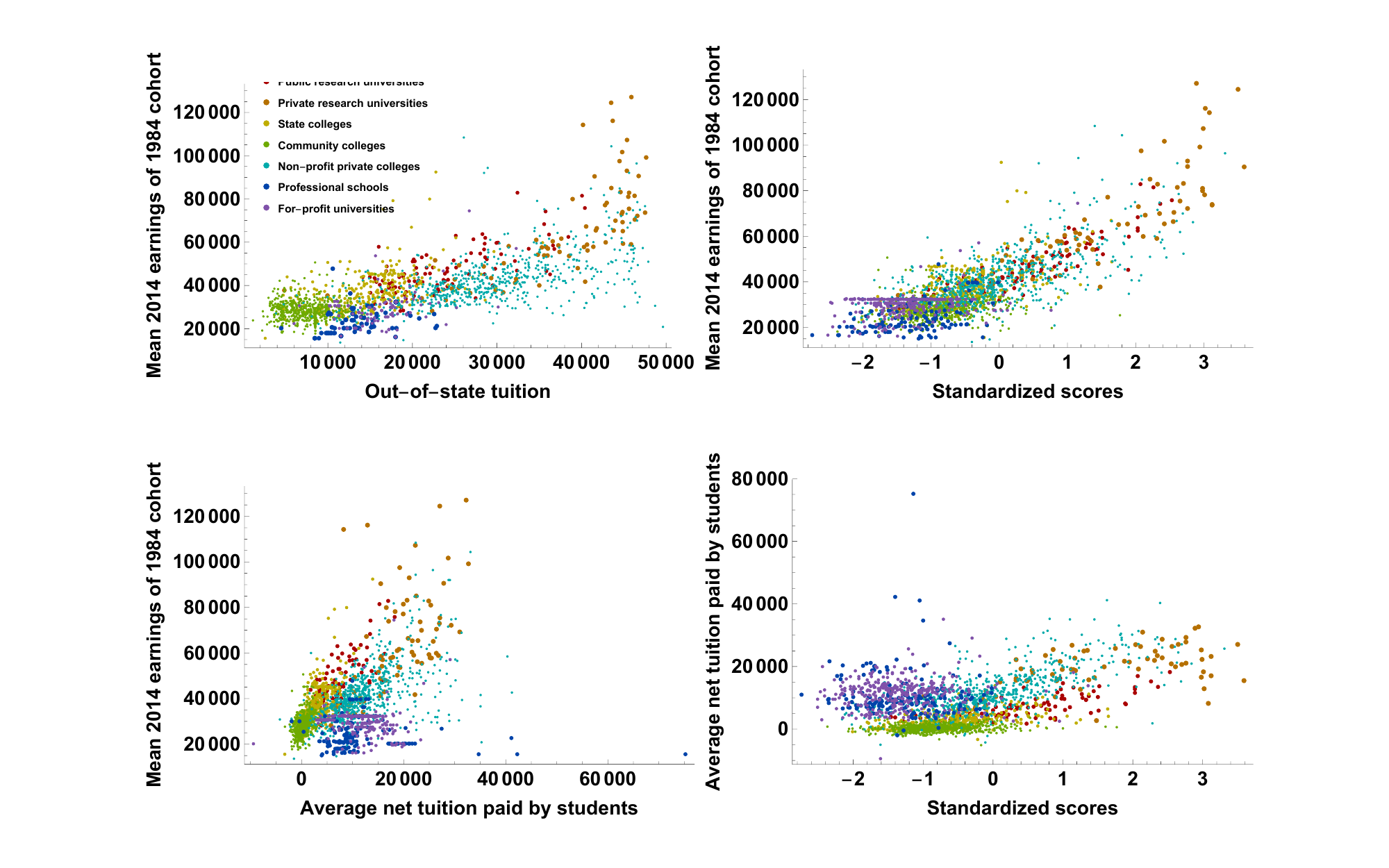}
\caption{Relationship between mid-career earnings of graduates and schools' tuitions and selectivity (See SI Figures H1-H2 for robustness of relationships to alternative choices of some of these variables).}
\label{fig:earnings}
\end{figure*}

\noindent {\bf Discussion}\\
Twenty-first century higher-education requires scalability. However, so far there has not been any mechanistic understanding for the tradeoffs and capabilities of universities both as a function of their institutional structure and educational mission, nor of their overall size.  Here, we have used scaling analysis to provide a synoptic view of the population of universities, which allows us to begin to characterize some important tradeoffs and capabilities of universities, provide a taxonomy of institutional scaling behaviors and assess the scalability of different sectors.

Our results display several interesting patterns that are common to all sectors. First, we see a split between sectors whose operations diversify with size (public and private research universities, and to some extent non-profit private colleges, which expand their student services) and those who specialize in teaching and exploit economies of scale in their instruction expenditure. Administration and maintenance scale sublinearly in all sectors (where we have the data), except in research universities, where they scale superlinearly with enrollment, but less steeply than productive functions, such as teaching or research expenditures. These findings are consistent with the structural theory of organization in sociology \cite{Blau1970}, where there are economies of scale for the administration of each operation, but an increase in administrative resources as operations diversify with the size of the organization.

Second, gains in efficiency with size are redeployed in ways that are, by and large, consistent with the core mission of these institutions. In research universities, increasing returns to size in the revenue from donations, research grants, and endowment correlate with a growth in research activities in larger schools. In state and community colleges, efficiences in teaching allow a dramatic fall in cost. In for-profit colleges, they allow greater profit margins. In professional schools and non-profit private colleges, they are redeployed towards student services, which is less obviously a core function.

The tradeoffs inferred from our scaling analysis across diverse sectors complement previous sociological and economic thinking on the U.S. higher educational system, which is more organizationally diverse than that of most other national systems of education. Largely a product of history \cite{GoldinKatz1998}, this diversity in form provides a diversity of function. It is uncontroversial to note that public universities provide key public goods - such as social cohesion through mass education, civic education, and research - which require subsidy. Non-profit private institutions enabled the professoriate to organize as a professional class, with autonomy from both the market and the state \cite{Menand2001}. Just as in the public form, the non-profit private form can also provide public goods, and may complement public institutions by providing a more differentiated service to sub-groups. The separation of research universities from teaching-only universities helps meet the greater demand for teaching than research in a system in which close to 70\% of high-school graduates attend some form of higher education, with community colleges playing a particularly important role in absorbing this demand for education \cite{KaneRouse1999}. For-profits are relative late-comers, and were originally focused on vocational training \cite{Demingetal2012}. They now provide generalist diplomas, but are still focused on professional training and cater to the growing share of working and adult students, who have substantially different needs than traditional students. 

Our analysis can help shed light on the tradeoffs students face as they consider schools of different size and in different sectors, and as they evaluate the odds of being admitted, the tuition costs and their expected earnings. Figure \ref{fig:earnings} combines the data on tuition with additional data on the mean SAT scores of incoming students and the mean earnings of students ten years after attending school, using the data assembled from tax returns by \cite{Chettyetal2017} (see SI Appendix H). Figures \ref{earningsvstuition} and \ref{scoresvsearnings} show that average mid-career earnings increase with a school's selectivity and its out-of-state tuition costs (i.e. the market price of attending the school without accounting for financial aid and state funding). These relationships are not surprising since being admitted to and graduating from more selective schools contributes in multiple ways to workplace success: 1) via signaling of ability, and 2) by learning from other high-ability peers and from well-paid faculty. In turn, selective schools can charge more in light of these high future expected earnings and, consequently, can spend more on students, further enhancing their future success. The fact that schools fall on a common curve suggests that schools from different sectors are competing in a common market to enroll students. While there is a common trend, we see that some sectors are highly clustered, and make a staged entry onto this ladder of educational cost and selectivity. The least selective schools are the for-profit private universities and professional schools, followed by state and community colleges. Non-profit private colleges span a large range of selectivity scores and earnings. Then come the public research universities, which are selective and hold a high earning potential, and finally the private research universities, which are the most selective, most expensive and generate the highest earnings. 

Of course, many students do not pay the full cost of out-of-state tuition. In-state tuition of public state schools is much lower than out-of-state tuition and students also receive fellowships and grants. Figure \ref{earningsvspaidtuition} shows average mid-career earnings as a function of the net tuition students actually pay on average, once state funding and grants are taken into account. The monotonic relationship of Figure \ref{earningsvstuition} remains, but with a much steeper slope: state funding and grants make a large difference to students' return on educational investment. Sectors clearly differ in the financial added value they provide. Public schools tend to offer a higher return on investment. At the high end, public and private research universities are indistinguishable. On the other end of the spectrum, professional schools and private for-profit schools cost a lot relative to other sectors given the low expected earnings they provide. Figure \ref{scoresvsearnings} shows that in part, this is related to their very low selectivity. Yet, at low selectivity, community colleges offer better prospects to students, costing very little and providing higher expected earnings than private for-profit and professionals schools \cite{Demingetal2012}. Non-profit private colleges (including liberal arts colleges) display extremely heterogeneous behavior, some of them rivaling the financial added value of research universities, while others resemble the behavior of for-profit private schools, and display high tuition costs relative to public schools with similar selectivity and earning potential.

One might wonder how certain sectors are able to maintain presence in the market when they offer lower earnings at the same tuition level. It would seem that the answer is that they provide educational options for students with lower scores, and it might be the case that these schools would struggle to compete if the public sector were to expand in size at the low-tuition and low-selectivity end of the spectrum.

We saw earlier that several sectors achieve substantial economies of scale in teaching, and that this is achieved by sublinear scaling in the number of faculty. In other words, sectors that achieve economies of scale in teaching do so by increasing average class sizes or faculty teaching load. Yet, this does not predict the scaling behavior of graduation rates, or of potential earnings, as shown in Table 2. Indeed, we see that sectors that display these deep cuts in teaching expenditure with size either display linear or superlinear graduation rates and linear average earnings, suggesting that class size or faculty burden are not fundamental factors in explaining the performance of universities and that economies of scale can be achieved in teaching without jeopardizing the performance of non-research schools. 

The organizational diversity of universities would suggest that no form is inherently more efficient and therefore diversity persists despite competition to enroll students. Our analysis nonetheless suggests that some sectors are more scalable than others. In particular, public research universities scale superlinearly in research funds and in student outcomes, while remaining more affordable than their private counterpart. In a complementary fashion, state colleges and community colleges offer drastic economies of scale. This translates into cost reductions to students or taxpayers while maintaining a constant or improving standard for student outcomes. Thus, to simultaneously optimize research output and access to education, our analysis suggests that investing in the growth of schools in these sectors is a valid strategy. In light of this, it is noteworthy that larger state and community colleges receive proportionally much less public funding. A future research goal is to identify how this has changed over time, and what the consequences have been for schools in these sectors and their ability to provide an affordable generalist education to a large number of students. Finally, it should be noted that there is significant variation around the central scaling relationships, which indicates that individual schools are able to achieve positive and negative shifts in performance at a given scale (see SI Figures D2-D11 and Tables D3-D4 for analysis of outliers). These outliers are indicative of institutions that may be experimenting with novel strategies and which deserve in-depth analysis in order to understand how internal streatgies lead to these deviations in outcomes and infer the constraints and options facing schools in a particular sector at a particular size.

\section*{Methods}
The original source of this data is the Integrated Postsecondary Education Data System, or IPEDS \cite{IPEDSDoc}, where we use the 2013 Delta Cost Project \cite{DeltaDoc2017} refinement of the IPEDS data. Spanning nearly the entire U.S. higher education system, it includes over 20 million students, from $5,800$+ accredited universities. We use total enrollment (undergraduate and graduate) as our measure of size (see SI Appendix C). We supplement this main data source with several other databases \cite{Chettyetal2017,ScorecardDoc} discussed below and in Appendix A, with the list and definitions of all variables in SI Table B1. 

We use completion data from two of the most-widely reported U.S. sources as measures of educational output. First, we use the IPEDS Graduation Rate Survey, included in the Delta dataset \cite{DeltaDoc2017}. This dataset tracks six-year completions for cohorts of first-time first-year degree-seeking students (FTFT) (see SI Appendix G). Second, we use student outcome data on cohorts of Federal Student Aid-receiving (FSA) students which is collected via FAFSA reporting and managed through the Department of Education's College Scorecard project \cite{ScorecardDoc}. The Department of Education considers these data usable for research, but excludes them from their consumer tool due to possible reporting inaccuracies (\cite{ScorecardDoc} p. 23-24). Both graduation rates describe cohorts that enrolled in 2007 and assess six-year outcomes by 2013 (excepting professional schools, where only a three-year rate was available). See SI Appendix E and G for details of the cross-dataset merging procedures and overall data limitations (specifically Table E1-E4 and G3 and Figures E1-E2 and G1-G2 on robustness of results to various aggregation problems). In particular, both FSA and FTFT cohorts used for our completion analysis can exclude or misrepresent  portions of the IPEDS total enrollment, and may therefore introduce error into our analysis of overall institutional performance. Here we favor FSA results, because we assume that aid-receiving cohorts are less prone to systematically misrepresenting the student body composition than traditional student cohorts.

For our analysis of mid-career earnings we rely on the data provided by the Mobility Reports Card project, part of the broader Equality of Opportunity project \cite{Chettyetal2017}. Data on incomes were obtained from tax filings and linked to individual students. The data that is made available is aggregated at the school level. We use the mean 2014 incomes of students who attended the school for at least one year, focusing on the cohort born in 1984.

\noindent {\bf Acknowledgments}\\
We gratefully acknowledge the support of the ASU-SFI Center for Biosocial Complex Systems and the National Science Foundation under Grant Number ACI-1757923. GBW would like to thank the Eugene and Clare Thaw Charitable Trust and the National Science Foundation under the grant PHY 1838420 for their generous support, and GBW and CPK thank CAF Canada for generous support.  ML would like to acknowledge support from the Smart Family Foundation, Volkswagen Foundation and the National Science Foundation SBE 1656284. ML and CPK would like to thank the Omidyar Fellowship at the Santa Fe Institute for Supporting this work. We thank Sidney Redner, Paul LePore and John Miller for valuable feedback. \\

\noindent {\bf Author Contributions}\\
CPK, MD, and GBW designed the concept of the study. RCT, XL, MD, and CPK performed the analysis. All authors discussed intermediate findings, interpreted results, and wrote the paper. \\

\noindent {\bf Competing Interest Statement}\\
The authors declare no competing interests.\\

\def\bibfont{\footnotesize}
\renewcommand{\bibsection}
{\noindent {\bf References}}

\bibliography{scalinguniversities}


\begin{table}
\begin{tabular}{| m{2cm} | m{1.6cm} | m{1.4cm} | m{1.5cm} | m{1.5cm} | m{1.5cm} | m{1.4cm} |}
\hline 
 Sector name & Control & Level & N (schools) & Avg enrollment & Sector enrollment & $\%$ Sector enrollment \\ 
 \hline 
 Public Research Universities & Public & 4yr+, Doc & 160 & 28,114 & 4,498,249 & 21.16 \\
 \hline
 Private Research Universities & Private & 4yr+, Doc & 102 & 11,656 & 1,188,915 & 5.59 \\
 \hline
 State Colleges & Public & 4yr+ & 382 & 9,569 & 3,655,440 & 17.19 \\
 \hline
 Community Colleges & Public & 2yr & 908 & 7,177 & 6,517,164 & 30.65 \\
 \hline
 Non-Profit Private Colleges & Private Non-Profit & 4yr+ & 1,373 & 1,839 & 2,524,604 & 11.87 \\
 \hline
  Professional Schools & For-Profit & 2yr, 2yr- & 2,230 & 312 & 695,753 & 3.27 \\
  \hline
 For-Profit Colleges & For-Profit & 4yr+ & 647 & 1,902 & 1,230,372 & 5.79 \\
 \hline
\end{tabular}
\caption{Description of sectors and their descriptive statistics. Doc: the school grants research doctoral degrees.}\label{allsectorstats}
\end{table}
\begin{table}
\begin{tabular}{|m{1.6cm}|m{1.6cm}|m{1.6cm}|m{1.6cm}|m{1.6cm}|m{1.75cm}|m{1.75cm}|m{1.75cm}|}
 \multicolumn{8}{c}{} \\
  \multicolumn{8}{c}{} \\
 \hline 
 \multirow{3}{*}{ Variable} & Public &  Private  &State& Community& Non-profit  & Professional & For-profit\\

 & Research & Research & Colleges & Colleges & Private & Schools &Colleges\\
& Universities &Universities &  &  & Colleges &  &\\ \hline
Teaching expenditure & $1.2 \pm 0.1$& $1.44\pm0.18$ &$0.9\pm0.04$& $0.81\pm0.02$& $0.99\pm0.02$& $0.97\pm0.03$ &$0.93\pm0.04$\\
 \hline
Tuition revenue & $1.18\pm0.09$ &$1.2\pm0.09$ &$1.04\pm0.06$ & $0.89\pm0.03$ & $1.15\pm0.02$ &$1.09\pm0.03$ & $0.99\pm0.03$ \\
\hline 
Research expenditure & $1.52 \pm0.31$ & $1.75\pm0.61$ & $0.89\pm0.27$ & - & - & - & - \\
\hline 
Research revenue & $1.29 \pm 0.23$ & $ 1.94\pm0.49$& $0.65\pm0.1$ & $0.71\pm0.05$ & $0.85 \pm 0.08$ & - & -\\
\hline 
Maintenance & $1.07\pm0.11$ & $1.33\pm0.18$ & $0.8\pm0.05$& $ 0.88\pm0.02$&$ 0.89\pm0.02$ & - &$ 0.75^*\pm0.15$ \\
\hline 
Total faculty & $1.16\pm0.09$&$1.18\pm0.14$&$0.88\pm0.04$&$ 0.84\pm0.02$&$ 0.89\pm0.02$&$ 0.76\pm0.02$&$ 0.83\pm0.04$\\
\hline 
Faculty pay & $1.2 \pm 0.1$ &$1.4 \pm 0.17$ & $0.91 \pm 0.04$ & $0.82 \pm 0.02$ & $0.98 \pm 0.02$ & - & $0.92^* \pm 0.24$ \\
\hline 
FSA completions & $1.09\pm0.07$&$ 1.09\pm0.09$&$1.11\pm0.05$&$ 1.\pm0.03$&$ 0.99\pm0.04$&$ 0.96^*\pm0.06$&$ 1.06^*\pm0.09$\\
\hline 
FTFT completions & $1.24\pm0.06$ & $ 1.17\pm0.04$ & $ 1.11\pm0.04$ & $ 0.79\pm0.04$ & $ 1.09\pm0.02$ & $1.02^*\pm0.02$ & $0.96^*\pm0.05$ \\
\hline 
Mid-career earnings& $1.09 \pm 0.11$ & $1.16 \pm .15$ & $1 \pm 0.03$ & $0.97 \pm 0.02$ & $1.18^* \pm 0.04$ & $0.96^* \pm 0.1$ & $0.95 \pm 0.05$\\
\hline
\end{tabular}
\caption{Scaling exponents for maintenance and production variables, as well as number and pay of personnel, and some measures of student outcomes as a function of university size. A blank space indicates absence of usable data, and a * indicates that over half of the universities in that sector are missing. FSA : students receiving Federal Student Aid. FTFT: First-time Full-time students. For all variables, university size is measured as total enrollment, except in the case of completions, where the scaling relationship is with respect to cohort size (see SI Tables G1-G2). See SI Figures I1-I14 for a complete analysis of the tradeoffs amongst sectors based on the scales at which features with different scaling exponents intersect. See SI Table H1 for sensitivity of mid-career earnings scaling to the year chosen to measure the size of the school. See SI Table J1 for a summary of all scaling results.}\label{outcome_exponents}
\end{table}

\end{document}